\documentclass[aps,twocolumn,showpacs,preprintnumbers,amsmath,amssymb,a4paper,superscriptaddress]{revtex4-1}

\usepackage{color}

\usepackage{colordvi}
\usepackage{amssymb}
\usepackage{amsmath}
\usepackage{epsf}
\usepackage{mathrsfs}
\usepackage{graphicx}
\usepackage{appendix}
\usepackage{hyperref}
\usepackage{float}
\usepackage[dvipsnames]{xcolor}

\begin{document}

\title{Symmetry breaking in binary Bose-Einstein condensates in the presence of an inhomogeneous artificial gauge field}

\author{S.~Sahar S.~Hejazi}
\email{sahar.hejazi@oist.jp}
\affiliation{Quantum Systems Unit, Okinawa Institute of Science and Technology Graduate University, Okinawa 904-0495, Japan}

\author{Juan Polo}
\affiliation{Quantum Systems Unit, Okinawa Institute of Science and Technology Graduate University, Okinawa 904-0495, Japan}

\author{Rashi Sachdeva}
\affiliation{Mathematical Physics and NanoLund, LTH, Lund University, Box 118, 22100 Lund, Sweden}

\author{Thomas Busch}
\affiliation{Quantum Systems Unit, Okinawa Institute of Science and Technology Graduate University, Okinawa 904-0495, Japan}

\date{\today}

\begin{abstract}
    We study a two component Bose-Einstein condensate in the presence of an inhomogeneous artificial gauge field. In response to this field, the condensate forms a localised vortex lattice structure that leads to a non-trivial symmetry breaking in the phase separated regime. The underlying physical mechanism can be understood by considering the energy landscape and we present a simplified model that is capable of reproducing the main features of the phase separation transition. The intuition gained by numerically solving this simplified model is then corroborated using the analytical Thomas-Fermi model.
\end{abstract}


\maketitle

\section{Introduction} \label{sec:Introduction}
 
Ultracold gases of neutral atoms have, in the past two decades, evolved into highly controllable systems that allow one to study and simulate numerous fundamental quantum mechanical effects \cite{Lewenstein:07,Bloch:08,Bloch:12}. One of the reasons for this is the large experimental toolbox for tuning almost all of the terms of their Hamiltonians using static or time-dependent external fields. This includes using spin-orbit coupling \cite{Zhang:16} or artificial gauge fields \cite{Dalibard:11,Goldman:14} to affect the kinetic part, optical lattices \cite{Bloch:08}, density-dependent gauge potential \cite{M_Edmonds2020_density_dep} or painted potentials \cite{Henderson:09} to adjust the external trapping terms, or Feshbach resonances to control the non-linear interaction terms \cite{Chin:10}, to name just a few. Additionally, systems with different symmetries can be created using multicomponent setups \cite{Taie:10,Capponi:16}.

The simplest multi-component system is a binary Bose-Einstein condensate (BEC) made either from bosonic atoms in two different hyperfine states \cite{Myatt:97}, two different isotopes \cite{Papp:08}, or two different elements \cite{McCarron_2011,Thalhammer_2008,Ferrari_2002,Modugno_2002}. These systems show intriguing physics related to interpenetrating superfluidity \cite{Fava_2018,Mingarelli_2019} and in particular possess a de-mixing phase transition \cite{Papp:08,Cipriani_2013,Nicklas_2011,Shrestha_2009}. The latter is mostly determined by the interplay between the different interaction energies and, in free space, occurs when the square of the inter-component interaction strength exceeds the product of the two intra-component interaction strengths. It is worth noting that for certain atomic condensate settings these three interaction strengths can in principle be tuned independently \cite{Thalhammer:08,Papp:08}. In non-homogeneous systems the point of the separation transition can be shifted, as the effects of the kinetic energy have to be taken into account \cite{Wen:12}. Other terms that can be present in the Hamiltonian, e.g.~accounting for Rabi coupling \cite{Merhasin:05}, spin-orbit coupling \cite{Gautam:14}, or rotations \cite{M_Edmonds2020_density_dep,Kasamatsu:03}, are known to have an influence on the phase separation threshold as well.

In this work we are interested in the phase separation process in a two-component system in the presence of angular momentum. However, contrary to previously considered situations \cite{Kasamatsu:18,Mingarelli_2019}, we will investigate systems where the rotational energy is not homogeneously distributed over the whole condensate. While such a situation can in principle be realised experimentally by locally creating vortices through phase imprinting \cite{Dobrek:99,ORiordan:16}, this technique usually leads to non-equilibrium situations as the condensate has to adjust its density to accommodate the imprinted phase distribution. To avoid excitations, such as phonon modes, which can have non-negligible effects on the phase-separation transition \cite{Lee_2018}, we instead consider spatially inhomogeneous artificial gauge fields that only induce rotation in certain areas of the condensate \cite{Lembessis:14,Mochol:15,Sachdeva:18}.

Condensates in harmonic traps have been shown to respond to homogeneous rotation with the formation of triangular vortex lattices \cite{Abo-Shaeer_2001}, whereas in different external potentials different geometrical arrangements of the vortices are possible \cite{PReijnders:04,McEndoo:09,LoGullo:11,Stockhofe:11}. Furthermore, condensates that encompass low-density regions, either due to local potential maxima or in the phase separated limit of a multicomponent system, can support the so called hidden or ghost vortices located in these regions
\cite{Kasamatsu:03,Kasamatsu:18,Tsubota:02,KasamatsuPRA:03,Wen:10,Wen:13}. While the response to localised rotation through a gauge field has already been explored for single-component condensates \cite{Lembessis:14,Mochol:15,Sachdeva:18}, the effect on the phase separation transition in two-component systems has not yet been discussed.

In order to clearly isolate the effects of localised rotation, we consider systems with as many symmetries as possible: both condensates are made from atoms of the same species, both have the same number of particles and both have identical intra-component interaction strengths. We also restrict ourselves to a fundamental two-dimensional dynamics and a rotationally isotropic trapping geometry. Without rotation the separation transition in such a system leads to a straight phase boundary that cuts through the center of the trapping potential and whose direction is due to spontaneous symmetry breaking.  In the presence of strong, global external rotation, this is no longer the case and the phase separation dynamics becomes highly complex and breaks all spatial symmetries by forming unordered {\it serpentine} vortex sheets \cite{Kasamatsu:03}. 

To create a situation which lies in between the non-rotating and globally rotating settings we consider a gauge field that originates from an evanescent optical field above the surface of a prism, close to which a two-component condensate is trapped. The short-range exponential decay of the evanescent field in the direction perpendicular to the prism surface then results in an artificial magnetic field with a pronounced maximum at some distance from the surface. While in the miscible regime this produces a localised vortex distribution in the direction parallel to the surface that is in principal consistent with the symmetric splitting of the two components, we show that the interplay between the kinetic and the interaction energy leads to additional symmetry breaking that is not purely determined by minimising the length of the phase boundary.

This manuscript is organised as follows. In Section \ref{sec:Model}, we describe our model for a two-component Bose-Einstein condensate in the presence of a non-homogeneous artificial gauge field originating from the evanescent field created at the surface of a dielectric prism. In Section \ref{sec:results} we show how this artificial gauge field affects the miscible and immiscible regimes and in Section \ref{sec:ToyModel} we study, through a simplified model, the physical mechanisms behind the symmetry breaking observed in the immiscible regime. The numerical results obtained are then corroborated using an analytical Thomas-Fermi model in Section \ref{sec:Thomas-Fermi symmetry breaking}, and finally we conclude in Section \ref{sec:summary and conclusions}.

\section{Model}\label{sec:Model}

\begin{figure}[tb]
\centering
   \includegraphics[width=\linewidth]{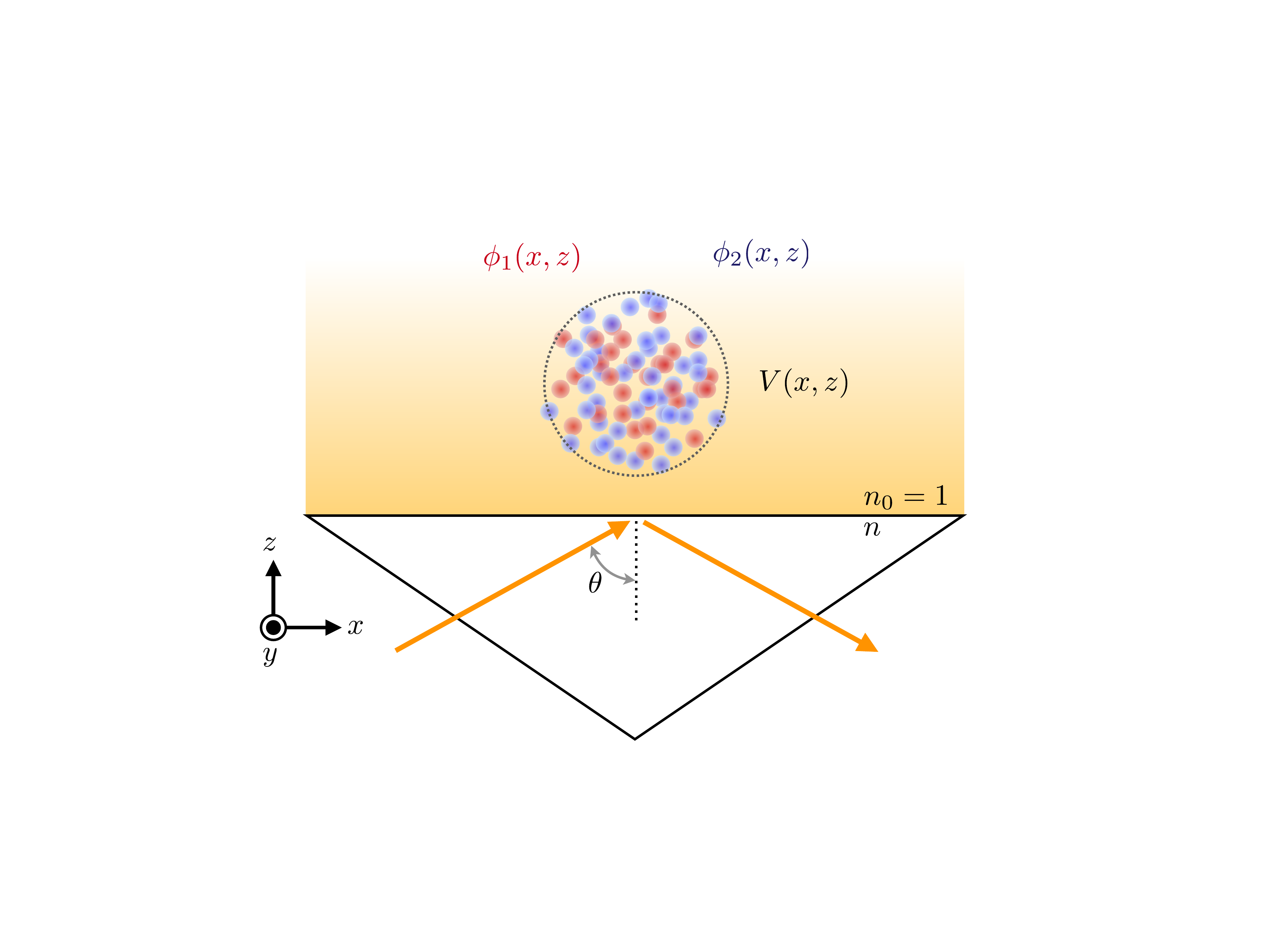}
   \caption{Schematic representation of a two-component BEC trapped in a potential $V(x,z)$ (geometry indicated by the dashed circle) just above the surface of a dielectric prism with the refractive index $n$. The center of both BECs is located at the origin of the coordinate system and and we always assume the surface of the prism to be at $z/a_{0}=-10$.}
   \label{fig:Schematic}
\end{figure}

We consider a two-component Bose-Einstein condensate of neutral alkali atoms that is tightly confined in one spatial direction, such that it can be effectively treated using a two-dimensional description. In particular, we choose a harmonic trapping potential of the form $V(x,z) = \frac{1}{2} M \omega^2 (x^2+z^2)$, where the frequency $\omega$ is the same in both directions, so that the trap is symmetric in the $x$-$z$ plane. Furthermore we assume that the atoms in both components have the same mass $M$, which can be achieved by trapping a single species and condensing the atoms in two different internal states. The entire system is located just above the surface of a dielectric prism with refractive index $n$, so that the atoms can interact with the evanescent field, see Fig.~\ref{fig:Schematic}.

Within the mean field approach, the two-component BEC can be described by a set of coupled Gross-Pitaevskii equations (GPEs) of the form \cite{Dalfovo:99}
\begin{align}
 i \hbar \frac{\partial \phi_ l}{\partial t} = \bigg[\frac{1}{2M} (P_l &- \mathbf{A})^2 + V (x,z)  \nonumber \\  
	&+ g_{ll} | \phi _l |^2 + g_{lm} |\phi _m|^2\bigg] \phi _l, \label{eq:GPE}   
\end{align}
where $g_{ll}=\frac{4 N \pi \hbar ^2 a_{l}}{M}$ and $g_{lm}=\frac{4 N \pi \hbar ^2 a_{lm}}{M}$ are the intra- and inter-component scattering strengths, respectively, with $l,m=\{1,2\}$ and $l \neq m$. As usual, $a_{l}$ is the s-wave scattering length between atoms of the same component and $a_{lm}$ for atoms of different components.  The condensate wave-function is described by $\phi_l$ and is normalized as $\int\!\! \int |\phi_l|^2 \, dx\, dz=1 $. The vector field $\mathbf{A}$ represents the gauge potential, which stems from the evanescent field emanating from the prism surface. 

To describe the gauge field we consider a laser field with a wave-vector $\mathbf{k}$ and frequency $\omega_{L}$, chosen to be close to the resonance of the atomic transition. This field propagates inside the prism at an angle $\theta$ with respect to its surface. When this angle is larger than the critical angle, $\theta_{0} = \arcsin (\frac{1}{n})$, the beam undergoes total internal reflection and an evanescent field is created at the surface of the prism. The electric field, $\mathbf{E} (x,z,t)$, propagates in $x$-$z$ plane with an amplitude $\mathbf{E}_{0}$ and decays from the surface in the positive $z$ direction with a penetration depth $d = (k_{0} \sqrt{n^{2} \sin ^{2} \theta -1} )^{-1}$. It takes the form
\begin{equation}
	\mathbf{E} (x,z,t) = t^\text{TE} (\theta) \mathbf{E}_{0} e^{- \mathrm{i} ( \omega_{L} t - \phi (x))} e^{-z/d} ,
	\label{E_evan}
\end{equation}
where $t^\text{TE} (\theta) = 2n \cos \theta \left( n \cos \theta + \mathrm{i} \sqrt{n^{2} \sin ^{2} \theta - 1 } \right) ^{-1}$ corresponds to the transmission coefficient, and the running phase is given by $\varphi (x) = x k_{0} n \sin \theta$ \cite{Mochol:15}.
 
The interaction between the evanescent field and the atoms in the condensate occurs via dipole coupling, $\mathbf{d} \cdot \mathbf{E} (x,z)$, where $\mathbf{d}$ is the electric dipole moment of the atoms. Without loss of generality we assume it to be the same for both components. In the rotating wave approximation this then leads to a dressed state of the form \cite{Mochol:15}
\begin{align}
  |\chi (x,z) \rangle =
    \left(
      \begin{array} {l}
          \cos [\Phi (x,z)/2] \\ 
          \sin [\Phi (x,z)/2] e^{- \mathrm{i} \varphi (x)} 
      \end{array} 
    \right),
 \end{align} 
where $\Phi (x,z) = \arctan \left( \frac{|\kappa(x,z)|}{\Delta} \right)$, $\kappa(x,z) = \mathbf{d} \cdot \mathbf{E} (x,z) / \hbar $ and $\Delta = \omega_{L} -\omega_{A}$ is the detuning of the laser light from the atomic resonance frequency, $\omega_A$, which we assume again to be the same for both components. 
Assuming that the atoms move slowly enough to adiabatically follow this spatially inhomogeneous eigenstate, they pick up a geometrical Berry phase which can be written as the appearance of a vector potential $\mathbf{A} =\mathrm{i} \hbar  \langle  \chi |\nabla \chi \rangle $, which has the explicit form
\begin{align}
	\mathbf{A} (x,z) &= \hbar \sin ^2 [\Phi (z)/2] \nabla \phi (x) \nonumber \\ 
				& =  \frac{n \hbar k_{0}}{2} \left[ 1- \frac{1}{\sqrt{1+ \left|\frac{\kappa(x,z)}{\Delta}\right|^2}} \right]   \sin \theta ~ \hat{\mathbf{x}}. 
\label{eq:AField}
\end{align} 
An artificial magnetic field can then be calculated from the vector potential via $\mathbf{B}  = \nabla \times \mathbf{A}$ as \cite{Mochol:15}
\begin{equation}
	\mathbf{B}(x,z) = - B_{0}\sqrt{n^2 \sin^2 \theta -1} \frac{s^2 \beta(z) n \sin \theta}{[1+s^2 \beta(z)]^{3/2}} \hat{\mathbf{y}}, 
	\label{eq:BField}
\end{equation}
\begin{figure}[tb]
\centering
   \includegraphics[width=\linewidth]{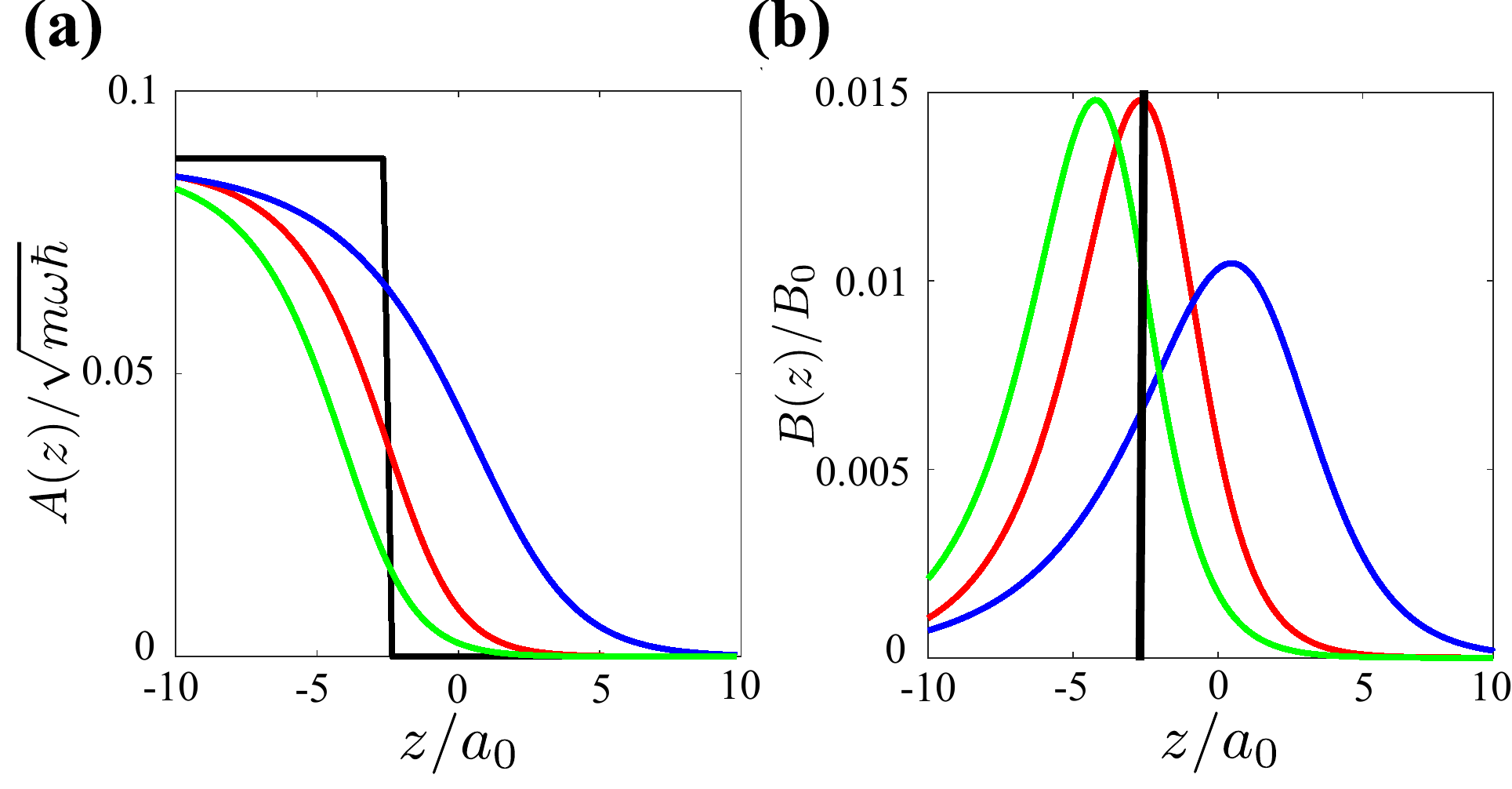}
   \caption{(a) Gauge potentials as a function of the distance above the prism surface. The green and red line correspond to an incident angle of $\theta-\theta_0 = 8 \times 10^{-4}$ rad, and $s=10$ and $s=20$, respectively. The blue curve corresponds to $\theta-\theta_0 =4 \times 10^{-4}$ rad for $s=20$. The black step function represents the model used in Sec.~\ref{sec:ToyModel}. (b) Normalized magnetic fields, $B(z)/B_0$, corresponding to the gauge potentials shown in (a) with the same colour coding, as plotted in \cite{Mochol:15}.}
   \label{fig:Schematic_AB}
\end{figure}

with $B_{0}= \hbar k_0 ^2/2 $, $\beta(z) = |t^{(\text{TE})}(\theta)|^2 e^{-2z/d}$ and $ s = \frac{| \mathbf{d} \cdot \mathbf{E}_{0}|}{\hbar | \Delta |}$. From Eqs.~\eqref{eq:AField} and \eqref{eq:BField} one can directly see that, since the evanescent field decays with increasing distance above the surface of the prism, the gauge field and the $B$-field will have to be inhomogeneous as well. For given sets of laser parameters this is shown in Fig.~\ref{fig:Schematic_AB}(a) for the $A$-field and in Fig.~\ref{fig:Schematic_AB}(b) for the $B$-field. In particular one can see from these plots that the artificial magnetic field has a maximum at finite distance away from the surface. The position of this maximum strongly depends on the value of $s$, while a change in the angle of the incident beam mostly affects the amplitude of the magnetic field. The atoms in a condensate trapped within the evanescent field will therefore experience effects corresponding to the presence of a spatially inhomogeneous $B$ field \cite{Lembessis:14,Mochol:15}. 

\section{Effects of the inhomogeneous magnetic field} \label{sec:results}

To explore the effects of the inhomogeneous artificial gauge potential, we numerically solve the two coupled Gross-Pitaevskii Equations~\eqref{eq:GPE} by using a standard FFT/split-operator method \cite{Feit_fft}. From here onward we work in harmonic oscillator units, that is $x\rightarrow x/a_0$, $z\rightarrow z/a_0$ and $t\rightarrow t\omega$,  with $a_0=\sqrt{\hbar/M\omega}$. We choose equal intra-component coupling coefficient, $g_{11} = g_{22} = g$ and inter-component coupling is given by $g_{12} = g_{21} \equiv \alpha g$. For stability reasons we only consider repulsive interactions. Thus, the condensate is in the miscible regime for $0<\alpha\lesssim 1$, and in the phase separated regime for $\alpha\gtrsim 1$. 

\begin{figure}[tb]
\centering
   \includegraphics[width=0.48\textwidth]{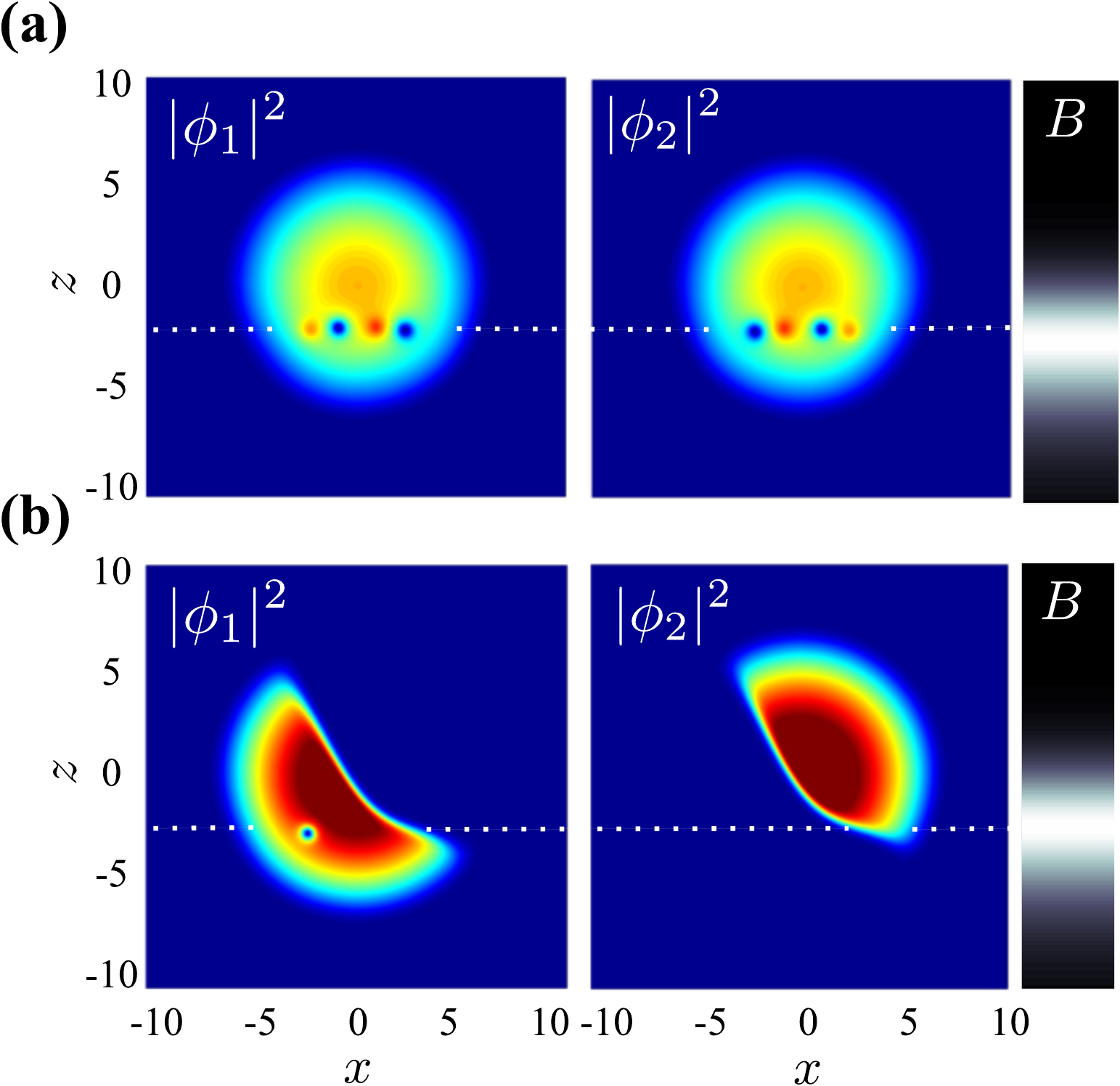}
   \caption{Ground state density profiles for the condensate trapped in the evanescent field in (a) the miscible ($\alpha =0.5$) and (b) immiscible ($\alpha =1.5$) regimes. We choose the parameter $s=20$ and a laser field that has an incident angle of $\theta-\theta_0 = 8 \times 10^{-4}$ rad with respect to the prism surface. The intensity of the $B$ field in the $x$ and $z$ direction is indicated on the right hand side. The artificial magnetic field is generated by choosing  $k_0 a_0= 1.0871\times10^{7}$ and $\kappa(x,z)/\Delta=20$.}
   \label{fig:Realistic}
\end{figure}

In Fig.~\ref{fig:Realistic} we show examples of ground state density profiles within the miscible and immiscible regimes. One can immediately note that the vortices only appear in a localised area, which corresponds to the region where the $B$-field is largest \cite{Lembessis:14,Mochol:15}. In the miscible regime and for the parameters chosen in Fig.~\ref{fig:Realistic} they form a single line along the maximum of the $B$-field, with each component having an offset with respect to the other such that they minimise the interaction energy. However, for less localised $B$-fields they can also arrange in a localised triangular lattices that converges to the full Abrikosov geometry for global fields \cite{Mochol:15}.

While in the immiscible regime the condensate components separate, as expected, it is immediately clear from Fig.~\ref{fig:Realistic}(b) that the separation can not be driven by the minimisation of the interaction energy alone. Naively one could expect that the separation boundary would be a straight line along the $z$-direction at $x=0$, which would lead to minimising the interaction energy and the kinetic energy stemming from the boundary, while ensuring that both condensates have the same amount of vortices and energy. However, this would not necessarily minimise the overall energy of the system, as additional kinetic energy is associated with the vortices. In fact, Figure~\ref{fig:Realistic}(b) shows that only one of the components carries visible vortices and that, even though the external parameters are the same for both situations, the number of vortices is not the same in the miscible and the immiscible regimes. This clearly indicates that some of the vorticity in the system is hidden in so-called ghost vortices, which are located in the low density areas at the phase boundary, so that the large rotational energy required to rotate high densities is avoided. To understand the interplay between the interaction and rotational energy in more detail, we will in the following explore a toy model of the inhomogeneous $B$-field that captures all relevant processes.

\section{Toy model}\label{sec:ToyModel}

Since the main characteristic of the inhomogeneous magnetic field is the existence of a localised maximum (see Fig.~\ref{fig:Schematic_AB}(b)), we will in the following consider the limit where the $B$-field is tightly localized in space. This can be achieved by assuming a step-like gauge-potential given by
\begin{equation}
    \mathbf{A}_{\Theta}(z)= A_0 ~ \Theta(-z + z_{0}),  
    \label{eq:A_field}
\end{equation}
where $\Theta(z)$ is the Heaviside step function, $z_{0}$ is the shift of the Heaviside function in the $z$ direction, and $A_0$ is the strength of the artificial gauge potential (see Fig.~\ref{fig:Schematic}(b)). The field is constant in the $x$-direction.
This form of the gauge potential catches the physical parameters that are related to the evanescent electric field created at the surface of the prism in a physically realistic and clean way: $A_0$ accounts for all the experimental parameters that characterize the strength of the realistic gauge potential (see Eq.~\ref{eq:AField}) and $z_0$ accounts for the shift in real space due to the $s$ parameter. In all our simulations below we chose $A_0$ such that its maximum value has the same order of magnitude as the realistic model.

\begin{figure}[tb]
\centering
   \includegraphics[width=0.48\textwidth]{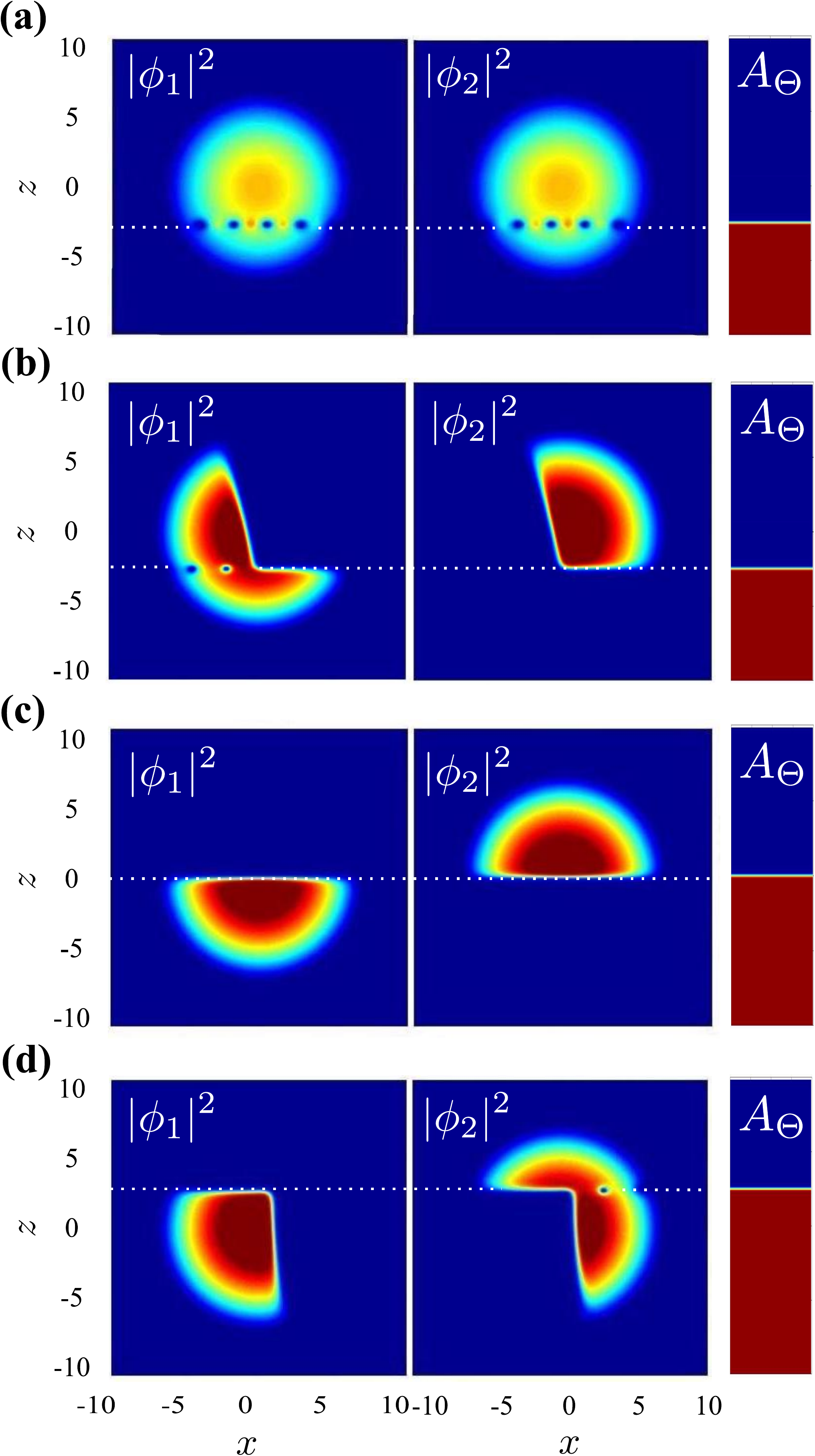}
   \caption{Ground state density profiles of each component of the two-component BEC in the presence of the step-function gauge potential (indicated on right hand side: red corresponds to finite value of $A_0=0.3$ and blue to zero). (a) Miscible regime with $\alpha=0.5$ and $z_{0} =-2.5$. (b-d) Immiscible regime with $\alpha =1.5$ and $z_0=-2.5$, $z_0=0$ and $z_0=2.5$, respectively. All density plots use the same color scale.} \label{fig:DenMiscIMisc}
\end{figure}

Typical ground state density distributions in the miscible and phase-separated regimes of the two-component system are shown in Fig.~\ref{fig:DenMiscIMisc}. In the miscible regime, $\alpha<1$, the localised $B$-field leads to a  single-line of vortices, see Fig.~\ref{fig:DenMiscIMisc}(a).  Due to the repulsive interactions between components, vortices within each component arrange themselves with an offset with respect to their counterparts, effectively filling the low density vortex cores of the other component. This is very similar to the realistic setting considered above, see Fig.~\ref{fig:Realistic}(a).

For the phase separation regime, $\alpha\gtrsim 1$, we show in Fig.~\ref{fig:DenMiscIMisc}(b-d) the density distribution for three different values of $z_0$. In panel (c) the $B$-field is located exactly at the center of the BEC ($z_0=0$), and one can see that this leads to a separation of the two components into two clouds with essentially mirror-symmetric density profiles. The phase boundary is exactly along the line of the finite $B$ field and corresponds to the shortest line possible. No vortices are visible and all vorticity is carried by ghost-vortices located in the low density area between the two components \cite{Kasamatsu:03,Kasamatsu:18,LoGullo:11}. This solution clearly minimizes the interaction and the kinetic energy of the system and is reminiscent of the standard phase separation in two-component systems without vorticity. However, the direction of symmetry breaking is now determined by the $B$-field and not chosen spontaneously.

For $z_0\neq 0$ this simple picture breaks down and the additional kinetic energy in the system plays a crucial role in how the phase separation occurs. In Fig.~\ref{fig:DenMiscIMisc}(b) and (d) we show the density profiles for $\alpha=1.5$ and the $B$-field located at $z_0=-2.5$ and $z_0=2.5$, respectively. 
One can immediately see that the two components separate in a non-symmetric way, which strongly depends on the position of the $B$-field, and that one component still possesses vortices, while the other does not. In particular one can see that the phase boundary is only partly along the line of the $B$ field, before turning to be more aligned along the $x$-direction. The part along the $z$-direction increases in length with increasing $z_0$, becomes the length of the whole condensate at $z_0=0$, and then decreases again almost symmetrically for $z_0>0$. While this symmetry breaking behaviour seems unusual at first sight, it can be intuitively understood by realising that the system is still trying to reduce the rotational energy by creating ghost vortices in the phase boundary region. Yet, when the $B$-field line does not cross the condensate symmetrically, separating the components fully along this line would lead to one component having a significantly smaller area available compared to the other. As the interaction energy is non-linear, this would lead to a significant increase in the overall energy, which is unfavourable. Thus, the system uses the $B$-field line partly to minimise the rotational energies, but then minimzes the interaction energies by departing from it. The phase separation is therefore a careful balance between the minimisation of the interaction and the rotational energies. It is important to realise that the situations for values of $z_0$ that are symmetric around zero are not fully identical (see Figs.~\ref{fig:DenMiscIMisc}(b) and (c)), as the $A$-field breaks the system's symmetry along $z$. Again, this is qualitatively the same behaviour that is also found in the realistic model shown in Fig.~\ref{fig:Realistic}(b).

\begin{figure}[tb]
\centering
   \includegraphics[width=0.5\textwidth]{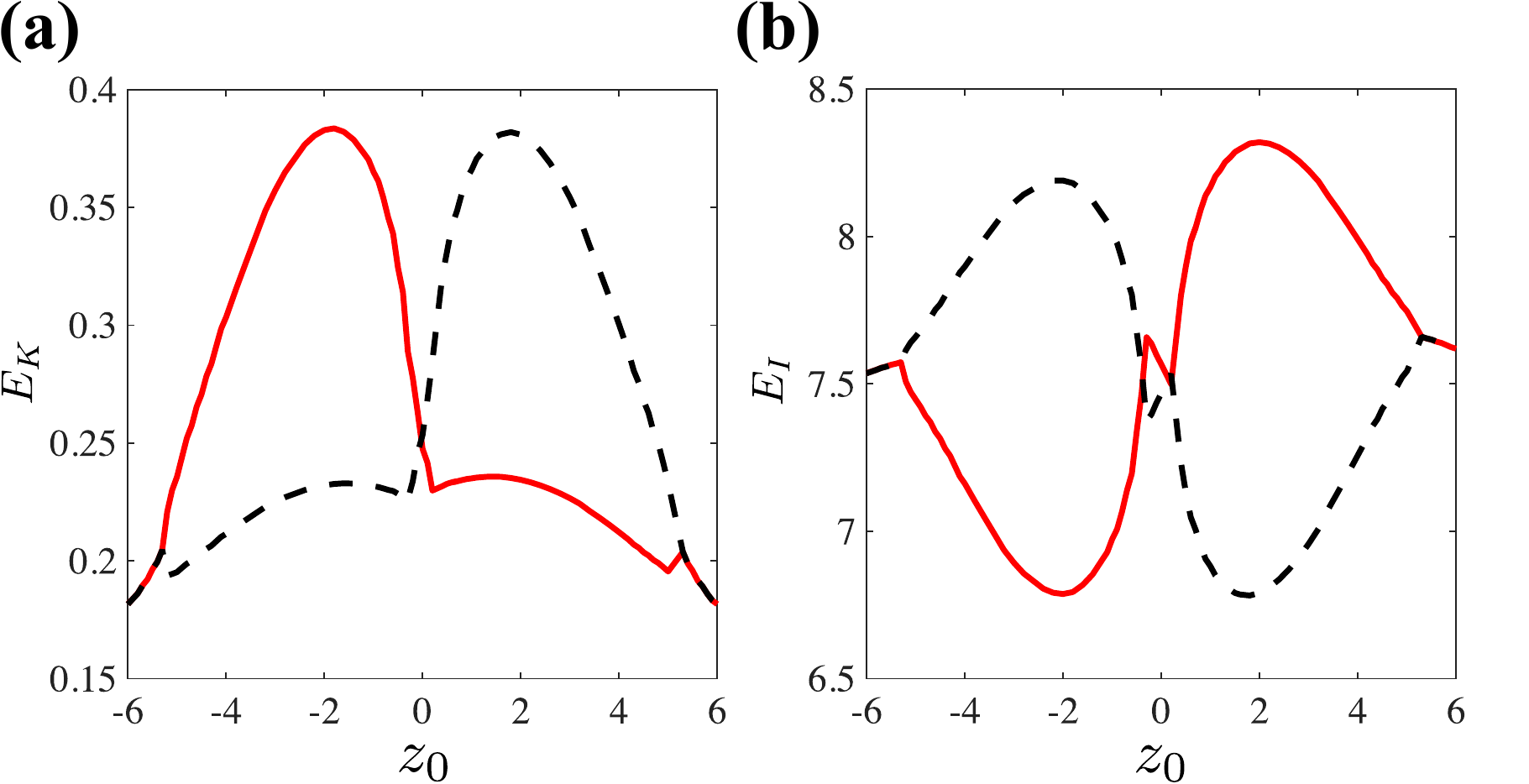}
   \caption{(a) Kinetic energy, $E_{\text{K}}^{(l)}$, and (b) interaction energy, $E_{\text{I}} ^{(l)}$, as a function of the position of the artificial magnetic field, $z_0$, for both components of the BEC in the immiscible regime ($\alpha=1.5$). The red (solid) line corresponds to the energy of first component and black (dashed) line to the energy of the second component. The artificial magnetic field is fixed at $A_0=0.3$ in harmonic oscillator units.} 
   \label{fig:EnergyComparison} 
\end{figure}
The intuition for the immiscible regime developed above can be supported by looking at the kinetic and interaction energy of each component given by
\begin{align} \nonumber
    E_\text{K}^{(l)}&=\frac{1}{2} \iint \phi_l^*( i \nabla_{l} + \mathbf{A}_\Theta )^2 \phi_l \; dx\, dz,\\
    E_\text{I}^{(l)}&= \frac{g}{2}\iint \left(|\phi_l|^4 + \alpha |\phi_m|^2 |\phi_l|^2\right) \; dx\, dz.
    \label{eq:energies}
\end{align}
One can see from Fig.~\ref{fig:EnergyComparison} that the kinetic energy of the vortex-carrying component grows initially much faster as the $B$-field line moves through the condensate, compared to that of the second component. However, with increasing values of $z_0$, the length of the phase separation border along the $B$-field line grows, leading to more and more vortices turning into ghost vortices. The kinetic energy therefore decreases again, until the same value is reached for both components when $z_0=0$, i.e. when all vortices have become ghost vortices. The same process then repeats in the second component, which starts carrying the vortices once $z_0>0$. 

The graph of the interaction energy as a function of $z_0$ (see Fig.~\ref{fig:EnergyComparison}(b)) shows that the component carrying the vortices has generally a lower non-linear energy than the one that carries no angular momentum. This is due to the additional centrifugal forces in the vortex carrying component, which allow the system to achieve a lower density. Again, these variations in energy go to zero when all vortices have been turned into ghost vortices at $z_0=0$ and the role of the two components flips subsequently. It is worth noting that when the $B$-field passes $z_0 =0$ a jump in the interaction energy can be seen  as there exist a sudden point when the last visible vortex has been turned into a ghost vortex. Again, the asymmetries present in the kinetic and interaction energies are due to the effect of the $A$-field, which increases the total energy of the system as it increasingly envelopes the entirety of the two component BEC. 

\section{Symmetry breaking}\label{sec:Thomas-Fermi symmetry breaking}

While the part of the phase separation line along the $B$-field line is set by external parameters, the remaining question is about the position of the turning point and the direction of the break away from it. Intuitively it should be as short as possible, which for rotationally isotropic geometries should lead to a break at a right angle.
One can see from Fig.~\ref{fig:DenMiscIMisc}(b)-(d) that this is approximately the case and below we confirm this intuition by determining the break-off point $(x_0,z_0)$ by energy minimization using the analytical Thomas-Fermi (TF) wavefunction, $\phi^\textrm{TF}_l$, obtained from solving 
\begin{align}
 \mu^\textrm{TF}_l \phi^\textrm{TF}_l = \bigg[V (x,z)  + g | \phi^\textrm{TF}_l |^2 + \alpha g |\phi^\textrm{TF}_m|^2\bigg] \phi^\textrm{TF}_l, \label{eq:TF} 
\end{align}
where $\mu^\textrm{TF}_l$ is the chemical potential of each component and we are again using harmonic oscillator units.
This approximation is valid when the kinetic energy terms of the Hamiltonian can be neglected as they are much smaller than the non-linear ones (see Fig.~\ref{fig:DenMiscIMisc}). 

\begin{figure}[tb]
   \includegraphics[width=0.48\textwidth]{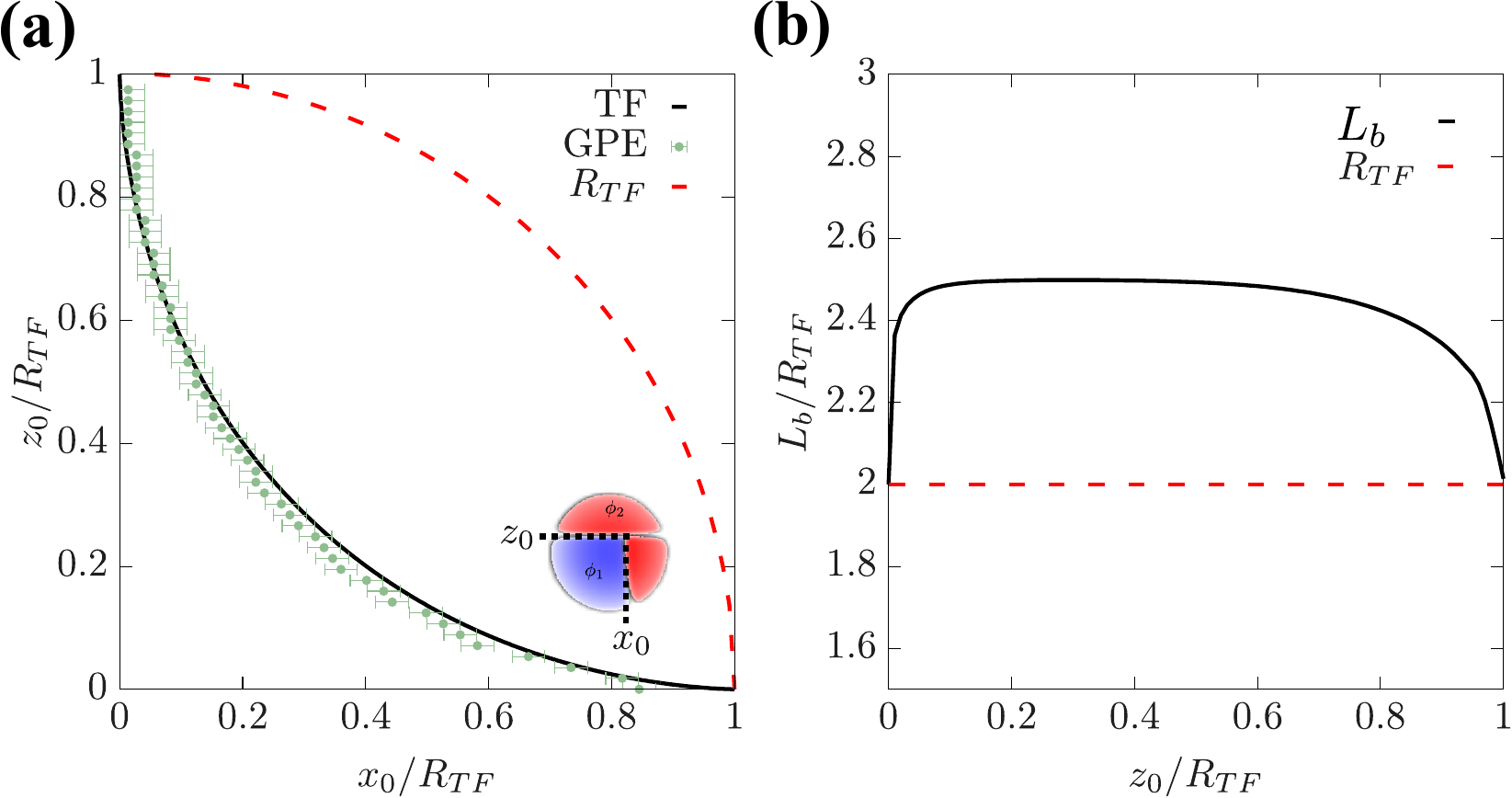}
   \caption{(a) Turning points $(x_0,z_0)$ of the
   asymmetric phase separation  (see Fig.~\ref{fig:DenMiscIMisc} for density plots) comparing the analytical Thomas-Fermi solution given by the solid-black line with the GPE numerical simulations, green dots (see \cite{Fitting} for details) for $\alpha=1.5$.  This plot shows the first quadrant of the 2D representation of the BEC, with $(0,0)$ corresponding to the center of the harmonic trap. The dashed-red line identifies the area delimited by the Thomas-Fermi radius found at $\alpha = 0$. (b)~Length of the phase boundary, $L_b$, within the TF approximation as a function of the position of the maxima of the magnetic field $z_0$. The dashed-red line indicates the shortest path for the phase boundary, i.e., $2R_{TF}$. }   
   \label{fig:thomas_fermi_vs_gpe} 
\end{figure}

However, as the kinetic energy clearly plays an important role in the phase separation, we take its effect into account by fixing the phase separation line along the maximum of the magnetic field in the $x$-direction up to a value of $x_0$. We then approximate the rest of the phase separation border by a straight line along the $z$-direction, so that both parts have a sharp $\pi/2$ angle between them (see inset of Fig.~\ref{fig:thomas_fermi_vs_gpe}). These conditions are encoded in the limits of integration for all integrals which depend on the position of the vertical part of the phase boundary, i.e.~$x_0$.
For simplicity we also fix the TF radius to $R_\textrm{TF}=\sqrt{2\sqrt{g/\pi}}$ in all limits of integration, so that no extra functional dependence on $x_0$ or $z_0$ appears. This allows us to avoid coupled transcendental equations for the chemical potentials of both components.

Within this model, we then minimize the total energy of the two-component system as a function of the position of the break away from the phase separation along the $B$-field line at $x_0$ (see \cite{Fitting} for details). The results from this analytical approach are shown in Fig.~\ref{fig:thomas_fermi_vs_gpe}(a) and one can immediately see that they very closely match the ones found from numerically solving the full two-component GPEs. This indicates that the straight line boundary along the $x$-direction connected at a right angle to the first part of the boundary along $z_0$  provides the lowest energy solution for the system to phase separate and forces it to spontaneously break the symmetry. We also show in Fig.~\ref{fig:thomas_fermi_vs_gpe}(b) that the length of the phase boundary exceeds the diameter of the condensate whenever the position of the $B$-field breaks the symmetry of the system. 
Finally, it is worth noting that while the assumption of a right angle connection between the two parts of the phase boundary is a good assumption in the toy model of the gauge field, it only approximately holds in the realistic model discussed in Sec.~\ref{sec:results}. This is due to the $B$-field being spread out over a larger range in the $z$-direction and the vortices being discreet along the $B$-field line. 

\section{Conclusions}\label{sec:summary and conclusions}

In this work we have investigated the ground state of a two-component Bose-Einstein condensate in the presence of a inhomogeneous artificial gauge potential. This situation appears when the gauge field is created by an evanescent field at the surface of a dielectric material, close to which the BEC is trapped, and it is therefore experimentally realistic. While in the miscible regime the systems responds to the angular momentum imposed by the gauge field in an expected manner, in the phase separated regime a careful balance between the need to minimise the interaction and the kinetic energy leads to a phase separation that spontaneously breaks the symmetry in unusual ways. In particular, the phase separation border is no longer just a straight line that crosses the system symmetrically as would be the case in the absence of the gauge field.

Using a toy model we have carefully explored the mechanism behind this symmetry breaking two-component state and clearly described the importance of the kinetic energy in the phase separation process. To confirm our numerical results, we have also presented an energy minimization calculation using the analytical TF solution that allows to determine the position where the phase boundary turns away from being a straight line.

Using inhomogeneous gauge potentials to induce rotation locally into condensates holds the potential to be a valuable way to engineer and study interesting superfluid dynamics. These can range from the above study on phase separation in multi-component condensates to creating well-defined initial states to study quantum turbulence \cite{schloss2019controlled}. The fact that such systems are experimentally possible using today's technology makes this an exciting and promising direction of research.

This project was supported by Okinawa Institute of Science and Technology Graduate University. JP also acknowledges the JSPS KAKENHI Grant Number 20K14417.

\bibliography{bibliography}

\begin{thebibliography}{47}
\bibitem [{Fit()}]{Fitting}%
  \BibitemOpen
  \href@noop {} {}\bibinfo {note} {The numerical data points are found by
  adjusting a Gaussian distribution onto the derivative of the density profile,
  which give us a mean value and variance for each fitting point,
  $\mu\pm\sigma$. We specifically calculate $\partial_z\phi_1(z)\approx
  \mathcal{N}(\mu,\sigma)$ for each $z_0$, with $\phi_1(z) = \int\,dx
  |\psi(x,z)|^2$ and $\mathcal{N}(\mu,\sigma)$ being the normal
  distribution.}\BibitemShut {Stop}%
\end{thebibliography}

\end{document}